\documentclass[pre,floatfix,showpacs,twocolumn]{revtex4}

\usepackage{amsmath}
\usepackage{amssymb}
\usepackage{graphicx}

\newcommand{\eg}{\emph{e.}$\,$\emph{g.}}
\newcommand{\ie}{\emph{i.}$\,$\emph{e.}}

\newcommand{\romd}{{\operatorname{d}}}

\newcommand{\VECe}{{\boldsymbol{e}}}
\newcommand{\VECf}{{\boldsymbol{f}}}
\newcommand{\VECl}{{\boldsymbol{l}}}
\newcommand{\VECn}{{\boldsymbol{n}}}

\newcommand{\VECt}{{\boldsymbol{t}}}

\newcommand{\VECX}{{\boldsymbol{X}}}

\newcommand{\VECsigma}{{\boldsymbol{\sigma}}}

\newcommand{\RR}{\mathbb{R}}

\newcommand{\CALC}{\mathcal{C}}
\newcommand{\CALH}{\mathcal{H}}
\newcommand{\CALL}{\mathcal{L}}
\newcommand{\CALS}{\mathcal{S}}

\begin{document}

\title{Contact lines for fluid surface adhesion}

\author{Markus Deserno}
\author{Martin M. M\"uller}
\affiliation{Max-Planck-Institut f\"ur Polymerforschung, %
             Ackermannweg 10, %
             55128 Mainz, %
             Germany}

\author{Jemal Guven}
\affiliation{Instituto de Ciencias Nucleares, %
             Universidad Nacional Aut\'onoma de M\'exico, %
             Apdo.\ Postal 70-543, %
             04510 M\'exico D.F., %
             Mexico}

\date{\today}

\begin{abstract}
When a fluid surface adheres to a substrate, the location of the
contact line adjusts in order to minimize the overall energy. This
adhesion balance implies boundary conditions which depend on the
characteristic surface deformation energies. We develop a general
geometrical framework within which these conditions can be
systematically derived.  We treat both adhesion to a rigid substrate
as well as adhesion between two fluid surfaces, and illustrate our
general results for several important Hamiltonians involving both
curvature and curvature gradients. Some of these have previously been
studied using very different techniques, others are to our knowledge
new.  What becomes clear in our approach is that, except for capillary
phenomena, these boundary conditions are \emph{not} the manifestation
of a local force balance, even if the concept of surface stress is
properly generalized. Hamiltonians containing higher order surface
derivatives are not just sensitive to boundary translations but also
notice changes in slope or even curvature.  Both the necessity and the
functional form of the corresponding additional contributions follow
readily from our treatment.
\end{abstract}

\pacs{87.16.Dg, 68.03.Cd, 02.40.Hw}

\maketitle


\section{Introduction}

There exist a number of physical systems whose energetics is fully
described by a surface Hamiltonian.  The easiest and best known
examples involve capillary phenomena
\cite{RowlinsonWidom,deGennesBrochardQuere}, where the shape of a
liquid-fluid interface (such as a sessile water droplet or the shape
of a fluid meniscus) is determined by minimizing its area. The same
physics governs the behavior of soap films.  Higher order surface
properties, notably its curvature, play a role in the description of
fluid lipid membranes or microemulsions \cite{Canham70,Helfrich73},
and even higher derivatives have been implicated in the occurrence
of certain corrugated membrane phases \cite{GoeHel96}.  In all these
cases the shape of the surface follows from minimizing the surface
Hamiltonian, a variational problem. The corresponding Euler-Lagrange
differential equations are known as the shape equations.

However, such surfaces are generally not isolated but rather in
contact with something else.  Water droplets or lipid membrane
vesicles may rest on a substrate, and this generally influences
their shape quite strongly.  For instance, water droplets on
hydrophilic substrates (\eg\ clean glass) resemble flat contact
lenses, while on very hydrophobic substrates (\eg\ teflon) they
are almost completely spherical.  When gravity can be neglected
\cite{gravity}, the shape equation dictates a constant mean
curvature surface in both cases (in fact, a spherical cap), but
the \emph{contact angle} at the three-phase line where water and
substrate meet is different for the two different substrates.

In the majority of cases the spatial extension of the surface
being studied exceeds the range of interaction between it and some
substrate by a large amount.  For instance, van der Waals forces,
hydrophobic interactions, or (screened) electrostatic forces
typically extend over several nanometers, while the extensions of
vesicles or droplets can be microns or even millimeters.  Under
these conditions the interaction is well approximated by a
\emph{contact energy}, \ie, an \emph{energy per unit area}, $w$,
liberated when the surface makes contact with the substrate.  It
is this adhesion energy, together with the energy parameters
characterizing the contacting surfaces, which determines the
boundary conditions holding at the contact line.  In the case of
capillary phenomena for example the ratio between adhesion energy
$w$ and surface tension $\sigma$ determines the contact angle
$\vartheta$ between liquid and substrate surface by means of the
well-known Young-Dupr\'{e} equation
\cite{RowlinsonWidom,deGennesBrochardQuere}
\begin{equation}
\frac{w}{\sigma}=1+\cos\vartheta \ . \label{eq:YoungDupre}
\end{equation}

The ``standard'' derivation of this result involves a balance of
tangential forces at the contact line.  Yet, despite being very
intuitive, this requirement of surface stress balance does not
yield the correct condition for more complicated surface
Hamiltonians, \emph{even} if the concept of surface stress is
generalized properly. Higher order Hamiltonians give rise to
additional energy contributions when the contact line is varied.
It is the purpose of the present article to show how these
contributions can be accounted for in a systematic and
parametrization-free way, and without assuming any additional
symmetries (such as axisymmetry or translation symmetry along the
contact line).  We study adhesion to rigid substrates as well as
to deformable surfaces also characterized by a surface
Hamiltonian. Our presentation generalizes and strongly simplifies
the analysis previously given in
Ref.~\cite{CapovillaGuven_adhesion}.

\section{Mathematical setup}

\subsection{Differential geometry}

In order to describe the adhering surfaces in a
parametrization-free way, we use a covariant differential
geometric language.  Our notation is essentially standard and will
follow the one used in
Refs.~\cite{CapovillaGuven_adhesion,surfacestresstensor,auxiliary,mem_inter_short,mem_inter_long}.
Briefly, a surface $\CALS$ is described by the embedding function
$\VECX(\xi^1,\xi^2)\in\RR^3$, where the $\xi^a$ $(a\in\{1,2\})$
are a suitable set of local coordinates on the surface.  This
induces two local tangent vectors
$\VECe_a=\partial\VECX/\partial\xi^a=\partial_a\VECX$ and a normal
vector satisfying $\VECn\cdot\VECe_a=0$ and $\VECn^2=1$.
Furthermore, the two fundamental forms of the surface are needed,
namely ($i$) the metric tensor $g_{ab}=\VECe_a\cdot\VECe_b$ and
($ii$) the extrinsic curvature tensor
$K_{ab}=\VECe_a\cdot\nabla_b\VECn=-\VECn\cdot\nabla_a\VECe_b$. The
symbol $\nabla_a$ is the metric-compatible covariant derivative.
The trace of the extrinsic curvature tensor will be denoted by
$K=K_a^a=K_{ab}g^{ab}$, which for a sphere of radius $r$ with
outward pointing normal vector is positive and has the value
$K=2/r$. As usual, indices are lowered or raised with the metric
or its inverse, respectively, and a repeated index $i$ (one up,
one down) implies a summation over $i=1,2$. More background on
differential geometry can be found in
Refs.~\cite{doCarmo,Spivak,Kreyszig}.

\subsection{Geometry at the contact line}\label{sec:geom_at_cl}

\begin{figure}
\includegraphics{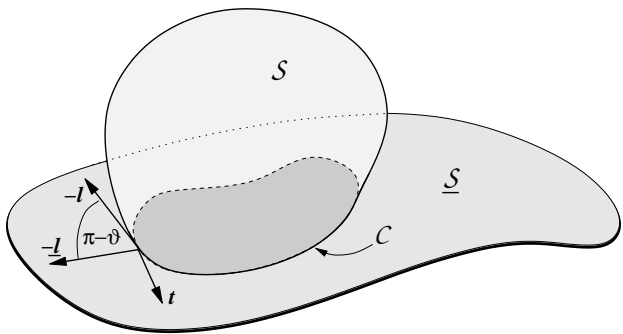}
\caption{Illustration of the geometry of surface adhesion.
Perpendicular to the contact line $\CALC$ two vectors exist,
$\VECl$ and $\underline{\VECl}$, which are tangential to surface
$\CALS$ and substrate $\underline{\CALS}$,
respectively.}\label{fig:general}
\end{figure}

By ``contact line'' we will denote the curve $\CALC$ at which the
surface detaches from the substrate.  Its local direction is given
by the tangent vector $\VECt=t^a\VECe_a$ (see
Fig.~\ref{fig:general}), which is tangential to $\CALC$, the surface
$\CALS$, and the substrate surface $\underline{\CALS}$ (which itself
might also be deformable). Here and in what follows (with the
exception of Sec.~\ref{ssec:adhesion_deformable}) we will use
underlining in order to indicate quantities referring to the
substrate.

Perpendicular to $\CALC$ we can define local normal vectors which
are either tangential to $\CALS$ or $\underline{\CALS}$, namely
$\VECl=l^a\VECe_a$ and $\underline{\VECl} = \underline{l}^a
\underline{\VECe}_a$, respectively (see Fig.~\ref{fig:general}).
Also, we will have two surface normals $\VECn$ and
$\underline{\VECn}$.  If the surface contacts the substrate at
zero contact angle, we will have $\VECe_a=\underline{\VECe}_a$,
$\VECn=\underline{\VECn}$, and $\VECl=\underline{\VECl}$ there;
however, their \emph{derivatives} perpendicular to the contact
line need not coincide, since the curvatures of surface and
substrate generally need not be identical.  In fact, the values of
perpendicular and parallel components of these curvatures (and
possibly their higher derivatives) will be among the primary focus
of this paper.  They will be denoted by
\begin{subequations}
\begin{alignat}{2}
K_{\perp} & \; = \; & l^a l^b K_{ab} & \; = \; -\VECn\cdot\nabla_{\perp}\VECl \ , \label{eq:K_perp} \\
K_{||} & \; = \; & t^a t^b K_{ab} & \; = \; -\VECn\cdot\nabla_{||}\VECt \ , \label{eq:K_par} \\
K_{\perp ||} \, = \, K_{||\perp} & \; = \; & l^a t^b K_{ab} & \; =
\; -\VECn\cdot\nabla_{\perp}\VECt \, = \,
-\VECn\cdot\nabla_{||}\VECl \ , \label{eq:K_perp_par}
\end{alignat}
\end{subequations}
where we also introduced the two directional surface derivatives
perpendicular and parallel to $\CALC$,
\begin{equation}
\nabla_\perp =l^a\nabla_a \qquad\text{and}\qquad
\nabla_{||}=t^a\nabla_a \ .
\end{equation}
Notice that we can analogously define $\underline{K}_\perp$,
$\underline{K}_{||}$, and $\underline{K}_{\perp ||}$.

\subsection{Hamiltonian}

In the present work we will exclusively study surfaces whose
energy $H$ is given by a surface integral over a scalar energy
density $\CALH$, which is constructed from the local surface
geometry:
\begin{equation}
H = \int_\CALS\romd A \; \CALH(g_{ab},K_{ab},\nabla_aK_{bc},
\ldots) \ .
\end{equation}
The integral extends over the entire surface $\CALS$, and the area
element $\romd A$ is given by $\sqrt{g} \, \romd\xi^1\romd\xi^2$,
where $g = \det(g_{ab})$ is the metric determinant.  For those
parts of the surface which adhere to a substrate we will assume an
additional adhesion energy density
\begin{equation}
\CALH_{\text{adhesion}}=-w(\xi^1,\xi^2) \le 0 \ ,
\end{equation}
which may in general be a function of position.

\section{Determining the boundary conditions}

\subsection{Continuity considerations}\label{ssec:continuity}

As we will see, the adhesion balance between surface and substrate
will result in a discontinuous change of some surface property
across the contact line.  However, the form of the energy density
restricts which quantities can be discontinuous, since it needs to
remain integrable.

Most obviously, the shape itself has to be continuous.  Yet, already
its first derivative may display a jump, as it does in the case of
capillary adhesion.  The energy density is given simply by
\begin{equation}
\CALH_{\text{capillary}}=\sigma \ , \label{eq:H_cap}
\end{equation}
and a kink in the surface at the contact line, \ie\ a finite contact
angle, is not associated with an extra energy.

For curvature elastic surfaces the situation is different. There,
the energy density of the surface is given by the well known
expression \cite{Canham70,Helfrich73}
\begin{equation}
\CALH_{\text{bend}} = \frac{1}{2}\kappa (K-K_0)^2 \ ,
\label{eq:H_bend}
\end{equation}
where $\kappa$ is the bending modulus (with units of energy) and
$K_0$ describes a spontaneous curvature of the elastic surface
\cite{GaussBonnet}.  A kink in the surface at the contact line
implies a $\delta$-singularity in the curvature, whose square is
non-integrable.  Hence, the surface needs to be differentiable
across the contact line, and the distinction between surface- and
substrate tangents and normals drawn in Sec.~\ref{sec:geom_at_cl}
becomes unnecessary.  Moreover, a quick glance at
Eqns.~(\ref{eq:K_par},\ref{eq:K_perp_par}) shows that both
$K_{||}$ and the off-diagonal curvature $K_{\perp ||}$ is
expressible as a tangential derivative along the curve $\CALC$ of
a quantity continuous across $\CALC$, hence both these curvatures
will also be continuous.  It is only the perpendicular curvature
component $K_\perp$ which might possess a discontinuity, and
indeed we will see that it does.

Finally, even higher order derivatives might occur in surface
Hamiltonians.  For instance, Goetz and Helfrich have studied a
curvature-gradient term of the form
\begin{equation}
\CALH_{\text{grad}} = \frac{1}{2}\kappa_\nabla (\nabla_a
K)(\nabla^a K) \equiv \frac{1}{2}\kappa_\nabla(\nabla K)^2 \ ,
\label{eq:H_grad}
\end{equation}
which in a generalized higher-curvature Hamiltonian prevents the
occurrence of infinitely sharp curvature changes \cite{GoeHel96}.
In this case it is obvious that all curvature components have to
be continuous along the substrate, since otherwise again a squared
$\delta$-singularity results.  Moreover, most of the first order
directional derivatives are automatically continuous:  The
parallel ones, $\nabla_{||}$, again differentiate quantities along
$\CALC$ which are continuous across $\CALC$ and thus are itself
continuous.  For the perpendicular ones it turns out that
$\nabla_\perp K_{||}$ and $\nabla_\perp K_{\perp ||}$ are
continuous, while $\nabla_\perp K_\perp$ is not.  This is
intuitively reasonable, since every term involving a ``$||$''
features at least one less derivative across the contact line and
thus cannot jump.  A rigorous proof is however a bit more
involved.  One may for instance proceed like this:  Start with the
contracted Codazzi-Mainardi equation
\cite{doCarmo,Spivak,Kreyszig} $\nabla_aK^{ab}-\nabla^bK=0$ and
project onto $l_b$.  By decomposing the resulting identity into
the local $(\VECl,\VECt)$ frame, it can be cast in the form
\begin{equation}
\nabla_\perp K_{||} = \nabla_{||}K_{\perp ||} +
(K_\perp-K_{||})\VECt\cdot\nabla_{||}\VECl+2K_{\perp
||}\VECl\cdot\nabla_\perp\VECt \ .
\end{equation}
Since every term on the right hand side is continuous across
$\CALC$ (recall that derivatives of tangent vectors are
essentially curvatures), $\nabla_\perp K_{||}$ must be continuous
as well. By projecting the contracted Codazzi-Mainardi equation on
$t_b$ instead, one can show continuity for $\nabla_\perp K_{\perp
||}$.

\subsection{Contact line variation}

For an adhering surface the total energy is stationary with respect
to variations of the contact line along the substrate. Such a
variation contributes twofold to the Hamiltonian:  Assume that
locally the contact line is moved such that a bit of surface unbinds
from the substrate.  This removes its corresponding binding energy,
as well as any elastic energy associated with the constraint of
conforming to the substrate, and thus gives rise to an energy change
$\delta H_{\text{bound}}$. On the other hand, the unbound part of
the surface acquires at the contact line a new boundary strip which
implies also a change $\delta H_{\text{free}}$ in its elastic
energy.  The boundary condition at the contact line then follows
from the stationarity condition
\begin{equation}
\delta H_{\text{cl}} = \delta H_{\text{bound}}+\delta
H_{\text{free}} = 0 \ . \label{eq:general_cl_variation}
\end{equation}
In the case of adhesion to a rigid substrate the bound
contribution involves the variation along a surface of known
shape.  The corresponding term is thus conceptually very different
from a deformable substrate or even the free variation, because in
both these cases the local shape of the surface is not known.
Below we will see how these differences manifest themselves when
computing the boundary terms.

\subsubsection{The bound variation}

For definiteness, let the normal $\underline{\VECl}$ to  the contact
line $\CALC$ be directed \emph{towards} the adhering portion of the
surface (see Fig.~\ref{fig:general}).  A local infinitesimal normal
displacement $\varepsilon$ of the contact line along a rigid
substrate thus implies the following obvious change in the bound
part of the surface:
\begin{equation}
\delta H_{\text{bound}} = -\int_\CALC \romd s \;
\big(\underline{\CALH} - w\big)\varepsilon(s) \ .
\label{eq:bound:variation}
\end{equation}
The underlining of $\CALH$ should again indicate that it is
evaluated with geometric surface scalars (such as for instance
curvatures) pertaining to the \emph{substrate}.  If the substrate
is flexible, the $w$ term remains, but the change in elastic
energy will instead be taken care of by an additional free
boundary variation.

\subsubsection{The free variation}

The change in energy due to the addition or removal of
\emph{unbound} parts to the boundaries of the surface is identical
to the boundary terms in the variation of the free surface. In
Ref.~\cite{auxiliary} it has been shown that for Hamiltonians up
to curvature order these terms are given by \cite{en_ne}
\begin{equation}
\delta H_{\text{free}} = -\int_\CALC \romd s \; l_a \Big[
\VECf^a\cdot\delta\VECX + \CALH^{ab}\VECn\cdot\delta\VECe_b\Big] \
. \label{eq:free_variation}
\end{equation}
Here, $\VECf^a$ is the surface stress tensor, given by
\begin{equation}
\VECf^{a} = (T^{ab}- \CALH^{ac} K_{c}^{b}) \VECe_{b}
      -(\nabla_{b} \CALH^{ab}) \VECn \ ,
  \label{eq:stresstensorcondeq}
\end{equation}
and we have also defined
\begin{subequations}
\begin{eqnarray}
\CALH^{ab} & = & \frac{\delta\CALH}{\delta K_{ab}} \quad\text{and}
\label{eq:Hab} \\
T^{ab} & = & -\frac{2}{\sqrt{g}}\frac{\delta
(\sqrt{g}\CALH)}{\delta g_{ab}} \ . \label{eq:Tab}
\end{eqnarray}
\label{eq:HabTab}
\end{subequations}
Finally, $\delta\VECX$ and $\delta\VECe_b$ denote the change of
contact line position and the associated change in the slope of
the tangent vectors, respectively.  Notice that the latter term is
only relevant if $\CALH^{ab}\ne 0$.

\section{Specific examples}

In this section we will illustrate the above formalism by applying
it to several important situations and surface Hamiltonians.  In
Sec.~\ref{ssec:adhesion_rigid} we first treat the problem of
adhesion to a rigid substrate.  We will see how known results (the
Young-Dupr\'{e} equation and the contact curvature condition for
Helfrich-membranes) follow with remarkable ease and can be extended
just as quickly to new Hamiltonians.  In
Sec.~\ref{ssec:adhesion_deformable} we look at the boundary
conditions involving adhesion to deformable substrates.
Specifically, in \ref{ssec:three_phase} we look at the triple line
between three tension surfaces, and in \ref{ssec:adhesion_vesicles}
we study the adhesion of two vesicles.

A central ingredient in all this will be the knowledge of the two
tensors $\CALH^{ab}$ and $T^{ab}$ defined in
Eqns.~(\ref{eq:Hab},\ref{eq:Tab}). While their determination is
not particularly involved, these calculations have been performed
previously by us
\cite{surfacestresstensor,auxiliary,mem_inter_short,mem_inter_long}
and we will thus simply reuse the results here.

\subsection{Adhesion to a rigid substrate}\label{ssec:adhesion_rigid}

Since the variation of the contact line has to proceed along the
substrate, we must have
\begin{equation}
\delta\VECX = \varepsilon \, \underline{\VECl} \ .
\label{eq:deltaX_substrate}
\end{equation}
No component in $\VECt$ direction is necessary, since for fluid
surfaces this would merely amount to a reparametrization of
$\CALC$. Notice that (\ref{eq:deltaX_substrate}) is nothing but
the Lie derivative of $\VECX$ along the substrate, since
$\CALL_{\varepsilon \underline{\VECl}}\VECX=\varepsilon
\underline{l}^a\nabla_a\VECX=\varepsilon
\underline{l}^a\underline{\VECe}_a=\varepsilon \,
\underline{\VECl}$. This property holds generally, and we will
make use of it later.

The normal component of the change in the surface tangent vectors
$\VECe_b$ only contributes if $\CALH^{ab}\ne 0$, \ie, if curvature
terms enter the Hamiltonian. We will assume that they do it in
such a way that differentiability of the surfaces is implied (see
Sec.~\ref{ssec:continuity}), so that no distinction needs to be
drawn between normal and tangent vectors of substrate and adhering
surface.  We then find
\begin{equation}
\VECn\cdot\delta\VECe_b=\VECn\cdot\nabla_b\delta\VECX=\varepsilon
l^c\VECn\cdot\nabla_b\underline{\VECe}_c = -\varepsilon
l^c\underline{K}_{bc} \ , \label{eq:deltaeb_substrate}
\end{equation}
where in the last step the equation of Weingarten
\cite{doCarmo,Spivak,Kreyszig} has been used; this is again the
Lie derivative along the substrate \cite{Lie_again}. Notice that
there still remains a distinction between \emph{curvatures} of
substrate and surface; hence, the derivative of the tangent
vectors resulting from a variation \emph{along the substrate}
yields the \emph{substrate} curvature and not the free surface
curvature.

\subsubsection{Capillary surfaces}

In this case the energy density is given by Eqn.~(\ref{eq:H_cap}),
and as we have seen in Sec.~\ref{ssec:continuity}, we will expect
a discontinuity in slope at $\CALC$. The bound variation is
\begin{equation}
\delta H_{\text{bound}} = -\int_\CALC \romd s \; (\sigma -
w)\varepsilon(s) \ .
\end{equation}
For this Hamiltonian we have
\cite{surfacestresstensor,auxiliary,mem_inter_short,mem_inter_long}
$\CALH^{ab}=0$ and $\VECf^a = -\sigma
g^{ab}\VECe_b=-\sigma\VECe^a$, and therefore
\begin{equation}
l_a\VECf^a\cdot\delta\VECX = -\sigma l_a \VECe^a \cdot \varepsilon
\underline{\VECl} = -\sigma\varepsilon \VECl\cdot\underline{\VECl}
= -\sigma\varepsilon\cos(\pi-\vartheta) \ ,
\end{equation}
where $\vartheta$ is the angle between capillary surface and
substrate -- in other words, the contact angle (see
Fig.~\ref{fig:general}). Equation (\ref{eq:general_cl_variation})
thus specializes to
\begin{equation}
\delta H_{\text{cl}} = -\int_\CALC\romd s \; \big[\sigma - w +
\sigma\cos\vartheta\big]\,\varepsilon(s) \ .
\end{equation}
Since $\varepsilon(s)$ is arbitrary, the term in square brackets
must vanish -- which gives the Young-Dupr\'{e} equation
(\ref{eq:YoungDupre}).

\subsubsection{Helfrich Hamiltonian}\label{ssec:Helfrich}

Let us now look at the energy density (\ref{eq:H_bend}) which
describes the continuum behavior of (tensionless) fluid lipid
bilayers. In this case $\CALH^{ab}=\kappa (K-K_0)g^{ab}$ and for
the stress tensor we have
\cite{surfacestresstensor,auxiliary,mem_inter_short,mem_inter_long}
\begin{equation}
l_a\VECf^a\cdot\VECl = -\frac{1}{2}\kappa(K-K_0)^2 +
\kappa(K-K_0)K_\perp \ .
\end{equation}
Together with the expressions for the contact line position and
tangent variation from
Eqns.~(\ref{eq:deltaX_substrate},\ref{eq:deltaeb_substrate}) we
find
\begin{subequations}
\begin{eqnarray}
\delta H_{\text{cl}} & = & -\int_\CALC\romd s \; \bigg\{ \frac{1}{2}\kappa(\underline{K}-K_0)^2 - w \nonumber \\
& & \qquad\qquad -\frac{1}{2}\kappa(K-K_0)^2 + \kappa(K-K_0)K_\perp \nonumber \\
& & \qquad\qquad -
\kappa(K-K_0)\underline{K}_\perp\bigg\}\;\varepsilon(s) \label{eq:Helfrich_cl_variation_bits} \\
& = & -\int_\CALC\romd s \; \bigg\{
\frac{1}{2}\kappa(K_\perp-\underline{K}_\perp)^2 - w
\bigg\}\;\varepsilon(s) \ , \label{eq:Helfrich_cl_variation}
\end{eqnarray}
\end{subequations}
where in the second step we made use of the continuity condition
$K_{||}=\underline{K}_{||}$ discussed in
Sec.~\ref{ssec:continuity}.  Equation
(\ref{eq:Helfrich_cl_variation}) implies a discontinuity in the
perpendicular curvature $K_\perp$ as the appropriate adhesion
boundary condition:
\begin{equation}
K_\perp - \underline{K}_\perp \; = \; \sqrt{\frac{2w}{\kappa}} \ .
\label{eq:contact_curvature_condition}
\end{equation}
The correct sign after taking the square root follows from the
fact that the detaching surface must not penetrate the substrate;
unfortunately this depends on ones specific choice of the surface
normal vectors.

Quite remarkably, this boundary condition depends neither on the
spontaneous curvature $K_0$, nor on the local parallel curvature
$K_{||}$.  It would also remain unaffected if the bilayer were
under a finite tension $\sigma$. Formally, it is easily seen to
cancel; physically, the reason is that the jump we would expect
from the Young-Dupr\'{e} equation (\ref{eq:YoungDupre}) cannot
materialize since the curvature terms in the energy density
enforce differentiability of the profile at $\CALC$.

Equation (\ref{eq:contact_curvature_condition}) coincides with the
result given previously in Ref.~\cite{CapovillaGuven_adhesion}. Its
axisymmetric version was first quoted in Ref.~\cite{Seifert90}, and
its specialization to a straight contact line can be found in
Ref.~\cite{LaLi_elast}.  We want to stress that the
$\CALH^{ab}\VECn\cdot\delta\VECe_b$ term in
Eqn.~(\ref{eq:free_variation}), which is responsible for the third
line in Eqn.~(\ref{eq:Helfrich_cl_variation_bits}), was crucial in
obtaining equation (\ref{eq:contact_curvature_condition}). Leaving
it out -- \ie, only treating the problem as a \emph{stress balance}
-- will \emph{not} result in the correct boundary condition.  The
only exception (treated via stress-balance in
Ref.~\cite{CapovillaGuven_adhesion}) is the special case of a flat
substrate, in which case $\delta\VECe_b\equiv 0$ and the missing
contribution vanishes anyway.

\subsubsection{General Hamiltonians involving $K$}

The analysis in the previous section readily extends to
Hamiltonians which are expressible in terms of the total extrinsic
curvature, $\CALH=\CALH(K)$.  One finds $\CALH^{ab}=\CALH'g^{ab}$,
where the prime denotes differentiation with respect to $K$, and
the tangential stress tensor contribution is
\begin{equation}
l_a\VECf^a\cdot\VECl = -\CALH + \CALH'K_\perp \ .
\end{equation}
The remarkably simple general boundary condition then reads
\cite{continuity}
\begin{equation}
(\underline{\CALH}-\CALH) - \CALH'(\underline{K}_\perp-K_\perp) \;
= \; w \ .
\end{equation}
Notice that unlike the form it takes for the quadratic Hamiltonian
discussed in Sec.~\ref{ssec:Helfrich}, this condition generally
involves the parallel curvature $K_{||}$ on the contact line.

\subsubsection{Curvature gradients}

Finally, we  want to study surfaces described by the Hamiltonian
density (\ref{eq:H_grad}), \ie, involving curvature
\emph{gradients}. From earlier work
\cite{mem_inter_short,mem_inter_long} we know that
\begin{subequations}
\begin{eqnarray}
\CALH^{ab} & = & -\kappa_\nabla(\Delta K)g^{ab} \qquad\text{and}\\
l_a\VECf^a\cdot\VECl & = & \kappa_\nabla \Big[ (\nabla_\perp K)^2
- \frac{1}{2}(\nabla K)^2 - K_\perp\Delta K \Big] \ . \;\;\;\;
\end{eqnarray}
\label{eq:Habf_grad}
\end{subequations}
However, simplemindedly inserting these expressions into the
formulas we have used so far does  \emph{not} give the correct
result.  Here is why: When calculating $\CALH^{ab}$ and $T^{ab}$
to obtain Eqns.~(\ref{eq:Habf_grad}), we varied the Hamiltonian
(\ref{eq:H_grad}) with respect to $K_{ab}$ and $g_{ab}$,
respectively, and identified the \emph{bulk} terms (see Sec. 4 in
the Appendix of Ref.~\cite{mem_inter_long}).  These variations
also leave \emph{boundary} terms, since the curvature
$K=g^{ab}K_{ab}$ appears differentiated.  For the purpose of
identifying $\CALH^{ab}$ and $T^{ab}$ they are irrelevant, but
they evidently matter now that we are interested in the total
energy change upon displacing the boundary.  Moreover, the
auxiliary framework introduced in Ref.~\cite{auxiliary} teaches us
that the $\delta K_{ab}$ and $\delta g_{ab}$ variations are indeed
independent from the $\delta\VECX$ and $\delta\VECe_a$ terms
already included in Eqn.~(\ref{eq:free_variation}), so they can be
simply added there. If we go through the calculation, we see that
the two tensor variations can be nicely combined into a single
scalar one, since they occur in the combination $g^{ab}\delta
K_{ab} + K_{ab} \delta g^{ab}=\delta K$, and we end up with the
additional boundary contribution
\begin{equation}
\delta H_{\text{free,grad}} = \kappa_\nabla \int_\CALC \romd s \;
l_a (\nabla^a K) \delta K \ . \label{eq:deltaK_variation}
\end{equation}
This is of course exactly the boundary term we would expect for
the variation of a Hamiltonian density whose functional form is
the square of the gradient of a scalar, so everything is
consistent.

To evaluate the right hand side of
Eqn.~(\ref{eq:deltaK_variation}), note that the variation is once
more given by the Lie-derivative along the substrate. Since $K$ is
a scalar, we obtain the simple expression
\begin{equation}
\delta K = \CALL_{\varepsilon \VECl} K = \varepsilon l^a\nabla_a
\underline{K} = \varepsilon\nabla_\perp \underline{K} \ .
\end{equation}
Together with Eqns.~(\ref{eq:Habf_grad}) we then obtain the total
contact line variation as
\begin{subequations}
\begin{eqnarray}
\delta H_{\text{cl}} & = & -\int_\CALC\romd s \;
\bigg\{ \frac{1}{2}\kappa_\nabla(\nabla\underline{K})^2 - w \nonumber \\
& & \qquad\qquad +\kappa_\nabla\Big[(\nabla_\perp K)^2 -
\frac{1}{2}(\nabla K)^2 - K_\perp\Delta K\Big]
\nonumber \\
& & \qquad\qquad +
\kappa_\nabla(\Delta K)\underline{K}_\perp \nonumber \\
& & \qquad\qquad -\kappa_\nabla (\nabla_\perp K)
(\nabla_\perp\underline{K})\bigg\}\;\varepsilon(s) \\
& = & -\int_\CALC\romd s \; \bigg\{
\frac{1}{2}\kappa_\nabla(\nabla_\perp
K_\perp-\nabla_\perp\underline{K}_\perp)^2 - w
\bigg\}\;\varepsilon(s) \ , \label{eq:grad_cl_variation}
\end{eqnarray}
\end{subequations}
where in the last step we used the continuity of curvatures and
their $\nabla_{||}$-derivatives (as discussed in
Sec.~\ref{ssec:continuity}) as well as the decomposition $(\nabla
K)^2 = (\nabla_\perp K)^2+(\nabla_{||}K)^2$. The boundary
condition following from this specifies a jump in the
perpendicular \emph{derivative} of the perpendicular curvature
\begin{equation}
\nabla_\perp \big(K_\perp-\underline{K}_\perp\big) \; = \;
\sqrt{\frac{2w}{\kappa_\nabla}} \ . \label{eq:grad_curv_condition}
\end{equation}
The similarity with Eqn.~(\ref{eq:contact_curvature_condition}) is
quite striking, and one might surmise a pattern that would be
followed by even higher derivative theories. Notice, however, that
the terms entering the derivation of
Eqn.~(\ref{eq:grad_curv_condition}) are quite different and that
the additional term stemming from
Eqn.~(\ref{eq:deltaK_variation}), which is absent in the simple
curvature square case, is essential.

\subsection{Adhesion to deformable surfaces}\label{ssec:adhesion_deformable}

\begin{figure}
\includegraphics{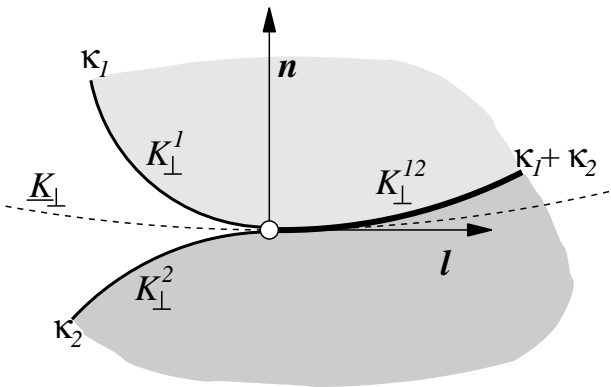}
\caption{Illustration of the geometry at the contact line between
two adhering vesicles.}\label{fig:vesicle_adhesion}
\end{figure}

Compared to the previous section, there are two key differences if
the substrate is not rigid.  First, the absence of a known
substrate shape along which a certain amount of deformation energy
is to be paid removes the term involving $\underline{\CALH}$ in
the bound variation (\ref{eq:bound:variation}).  Second, for the
same reason the contact line variation is no longer restricted to
proceed along a substrate and will thus be of the more general
form
\begin{equation}
\delta\VECX = \varepsilon_\perp\VECl+\varepsilon_n\VECn \ .
\label{eq:deltaX_general}
\end{equation}
The corresponding tangent vector variation, which occurs if
$\CALH^{ab}\ne 0$, then leaves a term
\begin{equation}
\VECn\cdot\delta\VECe_b =
\VECn\cdot\nabla_b(\varepsilon_\perp\VECl+\varepsilon_n\VECn)
 = -\underline{K}_{bc}l^c\varepsilon_\perp +
\nabla_b\varepsilon_n \ . \label{eq:deltaeb_general}
\end{equation}
Note that $\underline{K}_{bc}$ plays a different role here than
previously.  It no longer describes the curvature of the evidently
nonexistent substrate. Rather, the tangential variation may
proceed locally along a \emph{fictitious} surface which is
tangential to the other three surfaces that meet at the contact
line.  Encoding higher order derivative information necessary
here, $\underline{K}_{bc}$ describes the curvature of that
fictitious surface, and $\underline{K}_\perp$ is the component
perpendicular to $\CALC$ (see Fig.~\ref{fig:vesicle_adhesion}).
This surface is of course not unique, and thus
$\underline{K}_\perp$ is arbitrary -- just as the two variations
$\varepsilon_\perp$ and $\varepsilon_n$ themselves.

\subsubsection{Three phase capillary equilibrium}\label{ssec:three_phase}

The simplest example of a three phase line between deformable
surfaces occurs when three capillary interfaces meet, for instance
at the three-phase-line between three mutually immiscible liquids
1, 2, and 3, having mutual surface tensions $\sigma^{12}$,
$\sigma^{23}$ and $\sigma^{31}$. In this case no adhesion energy
is involved (or, alternatively, it may be considered as part of
the surface tension).  The contact line variation thus consists of
three identical boundary variations
\begin{eqnarray}
\delta H_{\text{cl}} & = & -\int\romd s \;
\Big\{l_a^{12}\VECf^{12a}+l_a^{23}\VECf^{23a}+l_a^{31}\VECf^{31a}\Big\}\delta
\VECX \nonumber \\
& = & \int\romd s \;
\Big\{\sigma^{12}\VECl^{12}+\sigma^{23}\VECl^{23}+\sigma^{31}\VECl^{31}\Big\}\delta\VECX
\ ,
\end{eqnarray}
from which we immediately find the boundary condition
\begin{equation}
\VECsigma^{12}+\VECsigma^{23}+\VECsigma^{31} = 0 \
.\label{eq:Neumann}
\end{equation}
This expresses nothing but the force balance between the three
directional line tensions $\VECsigma^{12}=\sigma^{12}\VECl^{12}$
etc.\ and is known as the \emph{Neumann triangle}
\cite{RowlinsonWidom}.  The vector equation (\ref{eq:Neumann})
corresponds to \emph{two} scalar equations (since there is no
component along $\VECt$). These are sufficient to determine the
three contact angles between the three phases (because their sum
equals $360^\circ$).  Notice that this conversely implies that by
measuring these angles one can only determine the \emph{ratios}
between the three tensions, not absolute values. How all this
information is conveniently extracted is discussed in detail in
Ref.~\cite[Chap. 8]{RowlinsonWidom}.

\subsubsection{Adhesion of two vesicles}\label{ssec:adhesion_vesicles}

For the case of two adhering vesicles we assume that vesicle 1 has
bending modulus $\kappa_1$ and tension $\sigma_1$, while vesicle 2
has corresponding values $\kappa_2$ and $\sigma_2$.  \emph{If} the
two bilayers can slide past each other in the region where they
adhere, their joint bending modulus is given by
$\kappa_{12}=\kappa_1+\kappa_2$, because the energies required to
bend either one just add; the same applies to the tension:
$\sigma_{12}=\sigma_1+\sigma_2$. We will for simplicity look at
the case where the spontaneous curvature is zero. The contact line
variation now contains one adhesion term and three free boundary
variations. Using the decomposition of $\delta\VECX$ and
$\delta\VECe_b$ as given in Eqns.~(\ref{eq:deltaX_general}) and
(\ref{eq:deltaeb_general}), respectively, we find the total energy
change to be
\begin{eqnarray}
\delta H_{\text{cl}} & = & -\int \romd s \bigg\{ \Big[ - w -
\frac{1}{2}(\kappa_1+\kappa_2)(K_\perp^{12})^2\nonumber \\
& &  +
\frac{1}{2}\kappa_1(K_\perp^1)^2+\frac{1}{2}\kappa_2(K_\perp^2)^2
\Big]\varepsilon_\perp \nonumber \\
& & - \nabla_\perp\Big[\kappa_1K_\perp^1+\kappa_2K_\perp^2
-(\kappa_1+\kappa_2)K_\perp^{12}\Big]\varepsilon_n \nonumber \\
& & -\Big[\kappa_1K_\perp^1+\kappa_2K_\perp^2-
(\kappa_1+\kappa_2)K_\perp^{12}\Big]\underline{K}_\perp\varepsilon_\perp
\nonumber \\
& & + \Big[\kappa_1K_\perp^1+\kappa_2K_\perp^2-
(\kappa_1+\kappa_2)K_\perp^{12}\Big]\nabla_\perp\varepsilon_n
\bigg\} \ . \;\;\;\;\;\;\;
\end{eqnarray}
All corresponding $K_{||}$ contributions cancel, since $K_{||}$ is
again continuous across $\CALC$; the same happens to the tensions.
The four terms belonging to the independent variations
$\varepsilon_\perp$, $\varepsilon_n$,
$\underline{K}_\perp\varepsilon_\perp$, and
$\nabla_\perp\varepsilon_n$ must vanish individually.  Since the
last two terms have identical prefactors, we finally end up with
three boundary conditions:
\begin{subequations}
\label{eq:v_adhesion_abc}
\begin{eqnarray}
\kappa_1(K_\perp^1)^2+\kappa_2(K_\perp^2)^2-(\kappa_1+\kappa_2)(K_\perp^{12})^2
& = & 2w \ , \;\;\;\;  \label{eq:v_adhesion_a} \\
\kappa_1K_\perp^1+\kappa_2K_\perp^2-
(\kappa_1+\kappa_2)K_\perp^{12} & = & 0 \ ,
\label{eq:v_adhesion_b} \\
\nabla_\perp\Big(\kappa_1K_\perp^1+\kappa_2K_\perp^2
-(\kappa_1+\kappa_2)K_\perp^{12}\Big) & = & 0 \ .
\label{eq:v_adhesion_c}
\end{eqnarray}
\end{subequations}
Notice that three conditions is just what we need in order to fix
three curvatures.  Yet, contrary to the case of vesicle adhesion
to a substrate, Eqn.~(\ref{eq:contact_curvature_condition}), these
conditions also contain one which involves \emph{derivatives} of
the curvatures, namely (\ref{eq:v_adhesion_c}). The origin of this
term stems from the $\varepsilon_n$-variation, which is forbidden
for the case of a rigid substrate.  Since the normal variation
also multiplies the normal component of the stress tensor, which
(as Eqn.~(\ref{eq:stresstensorcondeq}) informs us) always contains
one more derivative than the tangential one, this brings about the
higher derivative condition.

Regrettably, Eqns.~(\ref{eq:v_adhesion_abc}) do not look
particularly transparent.  It is however possible to symmetrize
Eqns.~(\ref{eq:v_adhesion_a},\ref{eq:v_adhesion_b}) and thus
obtain the two more suggestive equations
\begin{subequations}
\label{eq:v_adhesion_sym_ab}
\begin{eqnarray}
\Big(1+\frac{\kappa_1}{\kappa_2}\Big)\Big(K_\perp^1-K_\perp^{12}\Big)^2
& = & \frac{2w}{\kappa_1} \ , \label{eq:v_adhesion_sym_a} \\
\Big(1+\frac{\kappa_2}{\kappa_1}\Big)\Big(K_\perp^2-K_\perp^{12}\Big)^2
& = & \frac{2w}{\kappa_2} \ . \label{eq:v_adhesion_sym_b}
\end{eqnarray}
\end{subequations}
From Eqn.~(\ref{eq:v_adhesion_b}) it follows that one of the
$K_\perp^i$ is bigger and the other one smaller than $K_\perp^{12}$.
Hence, when taking the square root in
Eqns.~(\ref{eq:v_adhesion_sym_ab}), exactly one of the two will
necessitate a minus sign.

To conclude this section, let us look at two special cases of
these boundary conditions which turn out to be quite instructive.
First, if $\kappa_2\rightarrow\infty$, the second vesicle
approaches the limit of a rigid substrate.  In this case
Eqn.~(\ref{eq:v_adhesion_sym_b}) shows that
$K_\perp^{12}=K_\perp^2$ and Eqn.~(\ref{eq:v_adhesion_sym_a})
reduces to the old contact-curvature-condition we had just derived
for rigid substrates, Eqn.~(\ref{eq:contact_curvature_condition}).
And the curvature of this effective substrate is determined from
$\nabla_\perp(K_\perp^2-K_\perp^{12})=0$. This latter condition
shows that the ``substrate''-curvature is even differentiable
across $\CALC$ -- or, in other words, the ``substrate'' shape is a
three times continuously differentiable function.

And second, if the two membranes have identical bending moduli
$\kappa_1=\kappa_2=\kappa$, a ``symmetrized'' contact curvature
condition ensues which reads
\begin{equation}
\Big(K_\perp^1-K_\perp^{12}\Big)^2=\Big(K_\perp^2-K_\perp^{12}\Big)^2=\frac{w}{\kappa}
\ ,
\end{equation}
which tells us that the (squared) curvature jump $2w/\kappa$
demanded by the rigid substrate version
(\ref{eq:contact_curvature_condition}) is shared between the two
membranes, while the final condition becomes
$\nabla_\perp(K_\perp^1+K_\perp^2 -2K_\perp^{12})=0$.

\section{Summary}

We have shown how the boundary conditions pertaining to the
contact line between a fluid surface adhering to a solid substrate
or another deformable surface can be extracted from a systematic
boundary variation in a completely parametrization independent
way.  We would like to close with a summary of our main results
and some remarks:
\begin{itemize}
\item Integrability of the surface energy density $\CALH$ enforces
continuity of certain geometric variables across the contact line.
\item The highest derivative in $\CALH$ thus dictates which
geometric variables may change discontinuously across $\CALC$ in
response to adhesion. Hence, for the Helfrich Hamiltonian the
tension $\sigma$ does not enter the boundary condition even if it
enters $\CALH$; likewise, neither tension $\sigma$ nor bending
modulus $\kappa$ enter the boundary condition if also a
gradient-curvature-squared term is present in $\CALH$.
\item Higher order derivatives in $\CALH$ create boundary terms in
the variation which pick up surface variations that are one order
lower. If the curvature enters $\CALH$, then a change in slope is
noticed, if a gradient in curvature enters $\CALH$, then changes
in curvature are noticed.  For this reason the capillary
Hamiltonian is the \emph{only one} which only picks up
translations, such that the energy minimization can be
reinterpreted as a force balance. In all other cases higher
derivative deformations (such as torques or even more complicated
constructs) contribute to the boundary variation.
\item Generalizations to surfaces hosting additional scalar
or vector fields (such as composition or tilt order) appear
straightforward, since these are readily incorporated into the
present framework \cite{mem_inter_long}.
\item To be sure, knowing the boundary conditions does not mean
that one also knows the position of the contact line. Rather, the
latter has to be determined simultaneously with the surface shape.
In general this task is very difficult, but it is not the subject
of the present work.
\end{itemize}

\acknowledgments

We acknowledge the hospitality of IPAM, where this work was
originally conceived.  MD is grateful for the hospitality of UNAM,
where it was completed, as well as for an Emmy Noether grant
De775/1-3 by the German Science Foundation. JG acknowledges
partial support from CONACyT grant 51111 as well as DGAPA PAPIIT
grant IN119206-3.




\begin{thebibliography}{99}
\bibitem{gravity}
This is true for droplets small compared to the capillary length
$\ell=\sqrt{\sigma/\rho g}$, where $\sigma$ and $\rho$ are surface
tension and density of the liquid, and $g$ is the earth's
acceleration.  For instance, water under ambient conditions has
$\ell\simeq 2.6\,\text{mm}$.
\bibitem{GaussBonnet}
We have omitted a second quadratic term in the energy density
proportional to the Gaussian curvature, since by virtue of the
Gauss-Bonnet theorem \cite{doCarmo,Spivak,Kreyszig} it will not
contribute to the boundary conditions.
\bibitem{en_ne}
Equation (\ref{eq:free_variation}) is written slightly differently
in Ref.~\cite{auxiliary}.  That the two forms are identical may be
seen from the fact that since $\VECe_b\cdot\VECn=0$, we have
$\VECe_b\cdot\delta\VECn=-\VECn\cdot\delta\VECe_b$.
\bibitem{Lie_again}
This is most easily seen in the following way:
$\VECn\cdot\delta\VECe_b=-\VECe_b\cdot\delta\VECn =
-\VECe_b\cdot\CALL_{\varepsilon\VECl}\VECn =
-\VECe_b\cdot\varepsilon l^c\nabla_c\VECn =
-\VECe_b\cdot\varepsilon l^c \underline{K}_c^a\VECe_a = -
\varepsilon l^c g_{ab}\underline{K}_c^a = - \varepsilon l^c
\underline{K}_{bc}$, which is exactly the right hand side of
Eqn.~(\ref{eq:deltaeb_substrate}).
\bibitem{continuity}
For simplicity we have assumed differentiability of the profile
across $\CALC$, which holds (see Sec.~\ref{ssec:continuity}) if
$\CALH(K)$ is of higher than linear order.  The treatment of the
special case $\CALH\propto K$ can be found in
Ref.~\cite{CapovillaGuven_adhesion}.
\end{thebibliography}
\end{document}